# Fast PET Scan Tumor Segmentation using Superpixels, Principal Component Analysis and K-means Clustering

Yeman B. Hagos[1,*], Vu H. Minh[1], Saed Khawaldeh[1], Usama Pervaiz[1], Tajwar A. Aleef[1]

[1]Erasmus+ Joint Program in Medical Imaging and Applications, University of Burgundy (France), UNICLAM (Italy) and University of Girona (Spain)

* Email: yemanbrhane1989@gmail.com

*Abstract*—Positron Emission Tomography (PET) scan images are extensively used in radiotherapy planning, clinical diagnosis, assessment of growth and treatment of a tumor. These all rely on fidelity and speed of detection and delineation algorithm. Despite intensive research, segmentation remained a challenging problem due to the diverse image content, resolution, shape, and noise. This paper presents a fast positron emission tomography tumor segmentation method in which superpixels are extracted first from the input image. Principal component analysis is then applied on the superpixels and also on their average. Distance vector of each superpixel from the average is computed in principal components coordinate system. Finally, k-means clustering is applied on distance vector to recognize tumor and non-tumor superpixels. The proposed approach is implemented in MATLAB 2016 which resulted in an average Dice similarity of 84.2% on the dataset. Additionally, a very fast execution time was achieved as the number of superpixels and the size of distance vector on which clustering was done was very small compared to the number of raw pixels in dataset images.

## I. INTRODUCTION

Positron Emission Tomography (PET) is a non-invasive nuclear medicine functional imaging method that images the distribution of biologically targeted radiotracers with high sensitivity. PET imaging provides detailed quantitative information about many diseases and is often used to evaluate cancer with segmentation as a principal role. Image contrast enhancement is an essential pre-processing stage in image segmentation [1]. For several years, great effort has been devoted to the study of image enhancement techniques; wavelet-contourlet transform [2], iterative denoising and partial volume correction [3], iterative deconvolution [4] were few among them.

Segmentation can be thought as two consecutive processes, recognition and delineation. Recognition is determining where the targeted object is in the image, while the second process is defining the spatial extent of the recognized region [5]. [6], [7] demonstrated that manual segmentation is time-consuming, labor intensive, operator dependent, subjective, and these makes it less precise and reproducible. In the recognition process, regions of high uptake of tracer are identified either manually or automatically [8].

Although the number of PET image segmentation publications has always been lower than both CT and MRI [6], there have been some publications; graph cut and locally connected conditional random field via energy minimization [9], binary and Gaussian filtering regularized level set method with capability of detecting weak tumor boundary [10] were developed. In addition [11] developed k-means and fuzzy c-means clustering based segmentation; however, clustering was applied on image pixels directly and this in turns increases the execution time.

PCA based analysis of internal statistics of image patches gives tremendous insight to recognizing patterns in an image [12], which is applied to detect salient objects in natural images.

This paper presents implementation of unsupervised automatic PET image segmentation system to detect tumor regions from PET scans. Section 2 presents the mathematical formulation and implementation of the proposed approach which contains, contrast enhancement superpixels, PCA followed by k-means clustering to recognize the cancerous superpixels. Section 3 is devoted to discussion and evaluation of the simulation results. Finally, Section 4 concludes the paper.

## II. Implementation

The workflow of the proposed approach is divided into three stages: Preprocessing, Feature extraction, and clustering segmentation- where the second step can be divided into three sub-steps, and the third step into two as shown in Figure 1.

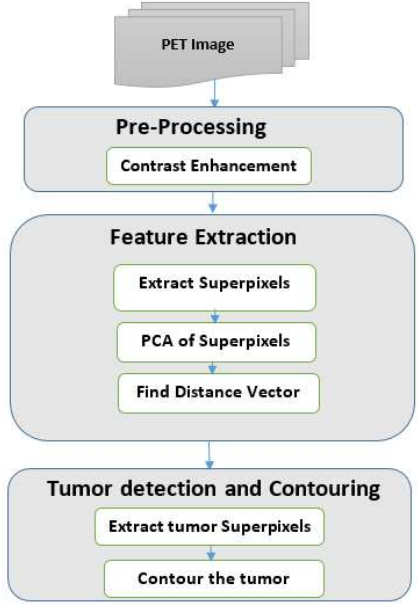

Figure 1: Implementation overview

### A. Preprocessing

Image enhancement is a subjective process to make the image suitable for the next step. In this paper, piecewise contrast enhancement was applied during the preprocessing part. Upon extensive observation from different images, it was found that by stretching pixels values greater than 110 to a range of gray values from 200 to 255 using piecewise linear stretching makes the image easy for clustering. This is mathematically shown in equation (1) below.

$$I_{enh} = \begin{cases} I, & \text{if } I \geq 110 \\ \frac{55}{145}(I - 110) + 200, & \text{otherwise} \end{cases} \quad (1)$$

where, I is input image and $I_{enh}$ is contrast enhanced image.

### B. Feature Extraction

Feature extraction is a process of simplifying the content of a large set of data in order to describe it efficiently for the purpose of facilitating further processing, storage requirement, and dimensionality reduction. In this paper, features are extracted using superpixel and Principal Component Analysis (PCA) as described below.

Superpixel is a group of pixels in proximity that has similar intensity. Simple Linear Iterative Clustering (SLIC) algorithm [13] is applied due to its fast computational time [14], [15]. The size of original superpixels extracted from SLIC is different as there might be a small number of pixels near each other with the similar pixel value in some of the region of the image(most of the time in tumor region), while the in non-tumor part of the image their size will be large. However, we need the same size of superpixels in order to apply PCA. This problem is solved as follows:

1) We computed average size of the superpixel as shown in equation (2).

$$M = \frac{1}{N} \sum_{i=1}^{N} n_i \quad (2)$$

Where N is the number of superpixels, $n_i$ is number of pixels in $i^{th}$ superpixel, M is average number of pixels per superpixel.

2) Then, the size of each superpixel is made same as of the average one by padding some pixel value to the smaller size superpixel and removing some intensity value from the large size superpixels. Instead of appending random intensity values to smaller size superpixels, we pad by repeating the last pixels value of the superpixel itself. Finally, the superpixel matrix is generated as shown in equation (3).

$$S = \begin{bmatrix} x_{11} & x_{12} & x_{13} & \cdots & x_{1N} \\ x_{21} & x_{22} & x_{23} & \cdots & x_{2N} \\ \vdots & \vdots & \vdots & \ddots & \vdots \\ x_{M1} & x_{M2} & x_{M3} & \cdots & x_{MN} \end{bmatrix} \quad (3)$$

Where each column represents a superpixel pixels, $M$ is in equation (2) and $N$ is number of superpixels.

As the goal is to detect pixels that are cancerous, and we know in PET images pixels that belong to the tumor have distinct intensity due to high uptake of radioactive tracer, so we need a method that analyses the internal statistics and makes an easy differentiation between the cancerous superpixels. PCA is one of the novel methods to study internal statistics of data. In addition to that, PCA reduces the dimensional space of the data [17]. In our implementation, PCA of superpixels is done as follow:

1) Compute average superpixel.

$$S_a = \frac{1}{N} \sum_{i=1}^{N} S_i \quad (4)$$

Where $S_i$ is the $i^th$ superpixel and $S_a$ average superpixel.

2) Determine the covariance of superpixels ($C_s$)

$$C_s = \frac{1}{N+1}(Y - Y_a^t)(Y - Y_a^t)^T \quad (5)$$

Where $Y$ superpixel matrix after average superpixel padding and $Y_a^t$ is mean of transpose of $Y$.

3) Calculate the eigenvectors and eigenvalues of the covariance matrix

$$C_s = P\Sigma P^T \quad (6)$$

Where $P$ is matrix with eigensuperpixels(principal components) as column and $\Sigma$ is diagonal matrix of eigenvalues.

4) Project the superpixels onto Principal components that contain most variance of the data. Here, the number of Principal components is same as the number of superpixels. As stated in [19], the eigenvectors or Principal components that contain at least 95% variance of superpixels can represent the whole image by confidence. This reduces the dimensional space as most of the information is contained in the first two or three largest eigenvalues. In our implementation, 95% variance of superpixels was contained in the top two principal components for most of the images. Once, the K dominant vectors are found for feature extraction (distance), the superpixel matrix is projected onto these dominant eigensuperpixels(eigenvectors) using equation (7).

$$Y_{Proj} = P_k^T Y \quad (7)$$

Where $P_k$ is eigenvectors matrix that contains at least 95% of the variation in the image and $P_{Proj}$ is the projection of superpixel matrix to $P_k$.

5) Calculate the distance of each superpixel in respect to average superpixel. While computing distance, we should consider the distribution of superpixels in the principal component coordinate system [12]. To incorporate this concept we computed the distance along the principal components. Mathematically, this will be computing $L_1$ norm distance in the principal components coordinate system as shown in Equation 8 below.

$$D(S_i) = \|\tilde{S}_i\|_1 \quad (8)$$

where $\tilde{S}_i$ is coordinate of $S_i$ relative to $S_a$ in the principal component coordinate system, and $D(S_i)$ is $L_1$ norm distance.

## C. Tumor Detection and Contouring

Currently, there are a variety of PET segmentation methods. The most commonly used methods are Fuzzy Locally Adaptive Bayesian (FLAB), Classification/Clustering, and some mixture of them. As stated in [6], there is a growing need for research in clustering based methods as they have the capability of detecting tumors with a complex shape in heterogeneous PET images. In our work, after distance vector is calculated in the principal components coordinate system, K-means clustering is applied. K-means is an algorithm that clusters a set of data based on distance measure. In our case, it separates the superpixels as tumor and non-tumor, which is binary classification using a minimization problem as shown in equation(9). Then, morphological operations(erosion and dilation) are then applied to delineate the spatial scope of the tumor.

$$\operatorname*{argmin}_{c} \sum_{i=1}^{k=2} \sum_{X \in c_i} D = \operatorname*{argmin}_{c} \sum_{i=1}^{k=2} \sum_{X \in c_i} \| X - \mu_i \|_2^2 \quad (9)$$

where $c_i$ is the set of points that belong to cluster i, $\mu_i$ is the center of $i^th$ cluster, X is distance vector extracted above and D is square of the Euclidean distance.

## III. RESULT AND DISCUSSION

Figure 2 shows the input image with a corresponding enhanced image in figure 3. It can be clearly seen that contrast between tumor and non-tumor region of the image is enhanced.

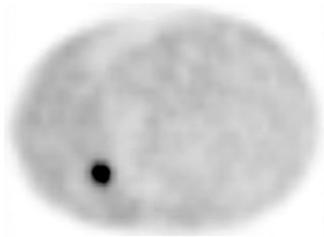

Figure 2: Input image

For the input image in Figure 2, it was found that 95% of the variance of superpixels is contained in the top two eigensuperpixels and Figure 4 presents scatter plot after the projection of superpixels onto the top two dominant eigensuperpixels. Principal component 1 is eigensuperpixel with the highest eigenvalue or the component that constitutes the highest variance of superpixel intensities. Principal component 2 is eigensuperpixel with the second highest eigenvalue. From the scatter plot it is evident that most of the superpixels are concentrated around the average superpixel (red star) as most parts of the image have similar pixel intensity distribution. The input image in figure 2 has 692 superpixels and as each of them are projected, there are 693 points in the scatter plot additional to average superpixel.

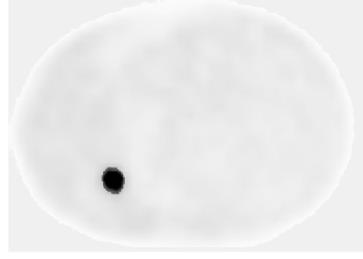

Figure 3: Enhanced image

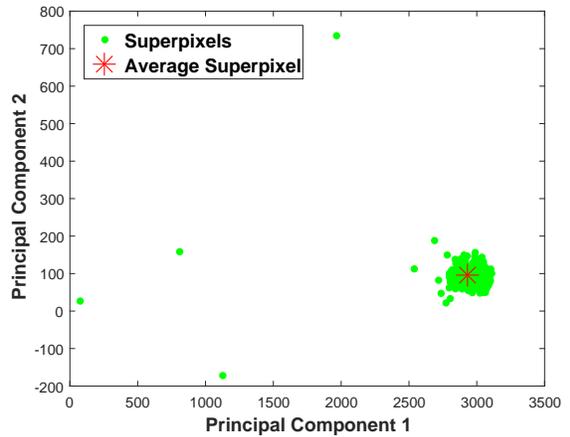

Figure 4: Scatter plot of projection of superpixels of the enhanced image onto the principal components

Figure 5 illustrates the $L_1$ norm distance of superpixels from their average along the principal components coordinate system. The horizontal axis represents the superpixel index and the vertical axis represents the distance from average superpixel. For the input image in Figure 2, the size of distance vector is 692 which far smaller than the size of the image (233x328). This is the reason why the execution time is so less for the proposed approach.

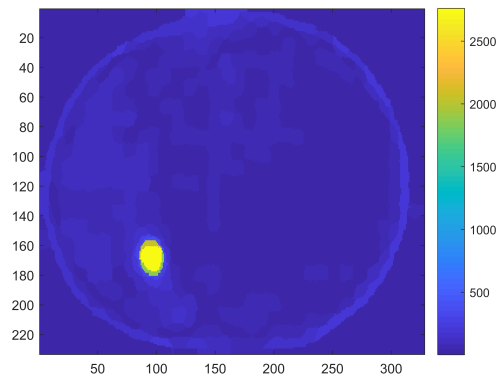

Figure 6: Heat map plot of superpixels distance in image space and superpixels K-means clustering

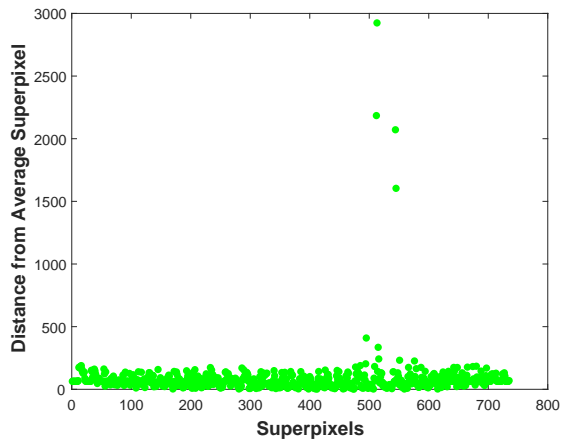

Figure 5: Distance of superpixels from average superpixel

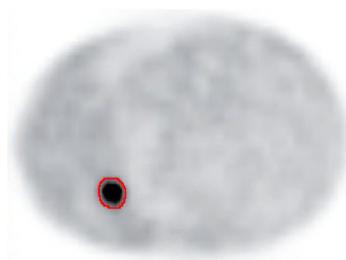

Figure 7: Output image with tumor contoured

Non-tumor superpixels (represented by a green point) are located near to average superpixel while tumor (red stars) are far from average. In addition, the heat map of the distance of superpixels from the average in the image space is shown in yellow color which is more distinguishable from the other superpixels with a large distance as depicted in the color bar. Internal statistics of tumor superpixels is so different from the average, thus, the distance will be very large. This classification can fail if cancerous part of the image is larger than the normal part. In case this happens, we have included another step to check some pixel values from each class so identification of the cluster to which the tumor belongs to can be more accurate.

Figure 7 shows the final result of our segmentation algorithm. As it is depicted in the figure the cancer region is identified and delineated correctly.

Table 1 contains information about the size of 3 sample images (2D) obtained from the 3 scans in the used dataset,

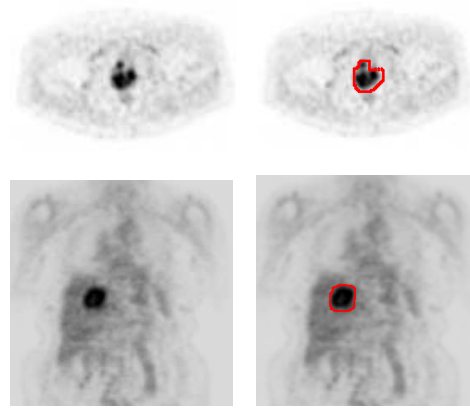

Figure 8: Input image and tumor segmentation results of some test PET images. The first column are input images with corresponding segmented image on second column.

number of superpixels, size of extracted features (distance), total execution time for this 3 sample, moreover, it shows the average segmentation Dice similarity of each complete scan. An example of segmentation result is shown in figure 8. Our implementation was tested on 100 images in total ( 45 from scan 1, 30 for scan 2, and 25 from scan 3). The size of extracted features and number of superpixels are smaller than the size of images in the dataset as shown in 1. The experiment was performed on an ordinary machine (i5-4210U CPU 1.7GHz). However, it provided a very short execution time of 2.35 ± 0.26 seconds for each image.

|  | Size | No superpixel | Distance vector | Execution time(s) | Average Dice Similarity(%) |
|---|---|---|---|---|---|
| Scan 1 Sample | 233 x 328 | 692 | 692 | 2.2 | 85.1 |
| Scan 2 Sample | 233 x 328 | 500 | 500 | 2.4 | 84.3 |
| Scan 3 Sample | 681 x 572 | 660 | 660 | 2.55 | 83.23 |

Table 1: Scan's sample sizes, number of supperpixels, size of distance vector, execution time and scan's average dice similarity.

Even that [16] and [18] were tackling a similar problem to the one presented in our work, however, they have not provided any measures of the execution time of their algorithms. The main concern of our paper was to design a fast PET tumor segmentation. As it can be seen from the table above, execution time of our proposed approach is very fast due to the following reasons: First, there is usually a small number of superpixels compared to the number of pixels in the image. Second, PCA again further reduces the dimension of the data which is then fed as the input to classification. In addition to that, MATLAB vectorization capability has been also extensively exploited throughout our implementation.

Additionally, Dice similarity for our algorithm was 84.2% which is very a comparable and competitive value with respect to the work in [16] and [18] as they obtained a Dice similarity measures of 80%-85% and 92%, respectively.

## IV. CONCLUSION

In this paper, we describe and evaluate PET image segmentation to extract cancerous part of the image. Piecewise contrast enhancement was first applied on the input image. Then, superpixel extraction and PCA was performed to extract feature for segmenting the image. After that, K-means clustering was applied to classify the image region into cancerous and non-cancerous parts. The experimental result shows that the proposed approach is capable of providing robust segmentation with fast execution time.

One of the major challenges encountered is the non-availability of public PET datasets to test the algorithm's performance on, that's why the algorithm was tested only on a small number of PET images. Therefore, testing and tuning the algorithm's parameters on other PET datasets surely will help increasing its generalization possibility.

ACKNOWLEDGEMENT(S)

Authors would like to thank Professor Alain Lalande and Professor Alexandre Cochet from University of Burgundy; Dijon (France) for providing the PET dataset for testing the implementation.